\documentclass[fleqn,10pt]{wlscirep}
\usepackage[utf8]{inputenc}
\usepackage[T1]{fontenc}
\usepackage{verbatim}
\usepackage{xcolor}
\usepackage{graphicx}

\newcommand{\suppref}[1]{\textit{#1}}
\title{Suppressing meta-holographic artifacts by laser coherence tuning}

\author[1,$\dag$]{Yaniv Eliezer}
\author[2,$\dag$]{Geyang Qu}
\author[2]{Wenhong Yang}
\author[2]{Yujie Wang}
\author[1,]{Hasan Yilmaz}
\author[2,*]{Shumin Xiao}
\author[2,*]{Qinghai Song}
\author[1,*]{Hui Cao}
\affil[1]{Department of Applied Physics, Yale University, New Haven, Connecticut 06520, USA}
\affil[2]{Ministry of Industry and Information Technology Key Lab of Micro-Nano Optoelectronic Information System, Shenzhen Graduate School, Harbin Institute of Technology, Shenzhen, 518055, China}

\affil[*]{E-mail: shumin.xiao@hit.edu.cn, qinghai.song@hit.edu.cn, hui.cao@yale.edu}

\affil[$\dag$]{These authors contributed equally to this work.}

\begin{abstract}
    A metasurface hologram combines fine spatial resolution and large viewing angles with a planar form factor and compact size. However, it suffers coherent artifacts originating from electromagnetic cross-talk between closely packed meta-atoms and fabrication defects of nanoscale features. Here, we introduce an efficient method to remove all artifacts by fine-tuning the spatial coherence of illumination. Our method is implemented with a degenerate cavity laser, which allows precise, continuous tuning of spatial coherence over a wide range with little variation in emission spectrum and total power. We find the optimal degree of spatial coherence to remove the coherent artifacts of a meta-hologram while maintaining the image sharpness. This work paves the way to compact and dynamical holographic display free of coherent defects.
\end{abstract}

\begin{document}

\flushbottom
\maketitle

\thispagestyle{empty}

\section{Introduction} \label{sec:introduction}

    Artificial Metasurfaces, comprised of a two-dimensional (2D) array of subwavelength scatterers, have shown unprecedented ability in controlling optical wavefront and converting conventional bulky optical elements into planar thin films \cite{chen2016review, glybovski2016metasurfaces, Genevet17}. One prominent example is the metasurface hologram (meta-hologram) \cite{Genevet_2015, kamali2018review, huang2018metasurface, Jiang19}. An ultrathin metasurface is capable of reconstructing a three-dimensional (3D) holographic image with a high spatial resolution and large viewing angles, while suppressing high-order diffraction \cite{ni2013metasurface, huang2013three, zheng2015metasurface, wan2017metasurface, lee2019recent}. Very recently, multi-color, multiplexed, and dynamic meta-holograms have been proposed and demonstrated, illustrating a great potential in information processing, 3D display, high-density data storage, and optical image encoding \cite{cui2014coding,chen2014high,wen2015helicity,segal2015controlling,gao2015broadband,ye2016spin,qu2020reprogrammable,fang2020orbital}. Despite of these remarkable advances, the road to practical applications of meta-holograms is hindered by coherent artifacts. Such artifacts originate from near-field interactions of subwavelength scatterers (meta-atoms), fabrication defects and phase dislocations and cause image distortion and degradation \cite{huang2018metasurface,yang2018freeform}.     
    While coherent artifacts and speckle noise are well-known issues for conventional holography, they are more significant in regard to meta-holography, as close packing of meta-atoms enhances the cross-talk and fabrication of nanoscale features is susceptible to error. Such artifacts cause severe distortions of holographic images, which are extremely difficult to correct. 

    One way of removing coherent artifacts is adjusting the coherence of an illumination source. For example, lowering the temporal coherence with broadband illumination provides spectral compounding \cite{McKechnie1975,Goodman10,GUO201641,deng2017coherence}. However, the meta-hologram is strongly dispersive, and the holographic image deteriorates with deviation from the design frequency. Another way is lowering the spatial coherence of illumination, which has been widely used for suppression of speckle noise in conventional holography \cite{bianco2018strategies}. It is done by either increasing the spatial coherence of a LED with spatial filtering \cite{dubois1999improved,deng2017coherence,lee2020light}, or decreasing the spatial coherence of a laser with moving elements and time integration \cite{shin2002viewing, Dubois:04, golan2009speckle, Takaki11}. While the former technique suffers from severe power loss, the latter requires long exposure time or computational costs. Moreover, reducing the spatial coherence will blur the image and reduce the depth of field, thus a precise tuning of the spatial coherence is required. 

    To eliminate coherent artifacts of meta-holograms, we resort to a degenerate cavity laser (DCL) with tunable spatial coherence for illumination. The DCL provides a wide tuning range of spatial coherence with little power loss \cite{Nixon13,Knitter16,Liew17}. Its fast decoherence enables a short exposure time for high-speed imaging \cite{Cao19}. Furthermore, the emission spectrum of the DCL does not change during the spatial coherence tuning, avoiding the spectral dispersion of the meta-hologram. By fine-tuning the spatial coherence of the DCL, we find the optimal degree that removes all sorts of artifacts without a significant blurring of the holographic image. Our scheme works efficiently for different types of meta-holograms, providing a general method for artifact-free holographic display. 

\section{Coherent artifacts created by meta-holograms} \label{sec:speckle}

\begin{figure}[htbp]
    	\centering
    	%\fbox{
    	\includegraphics[clip, trim = 0.5cm 1cm 3cm 0cm, width=1\linewidth]{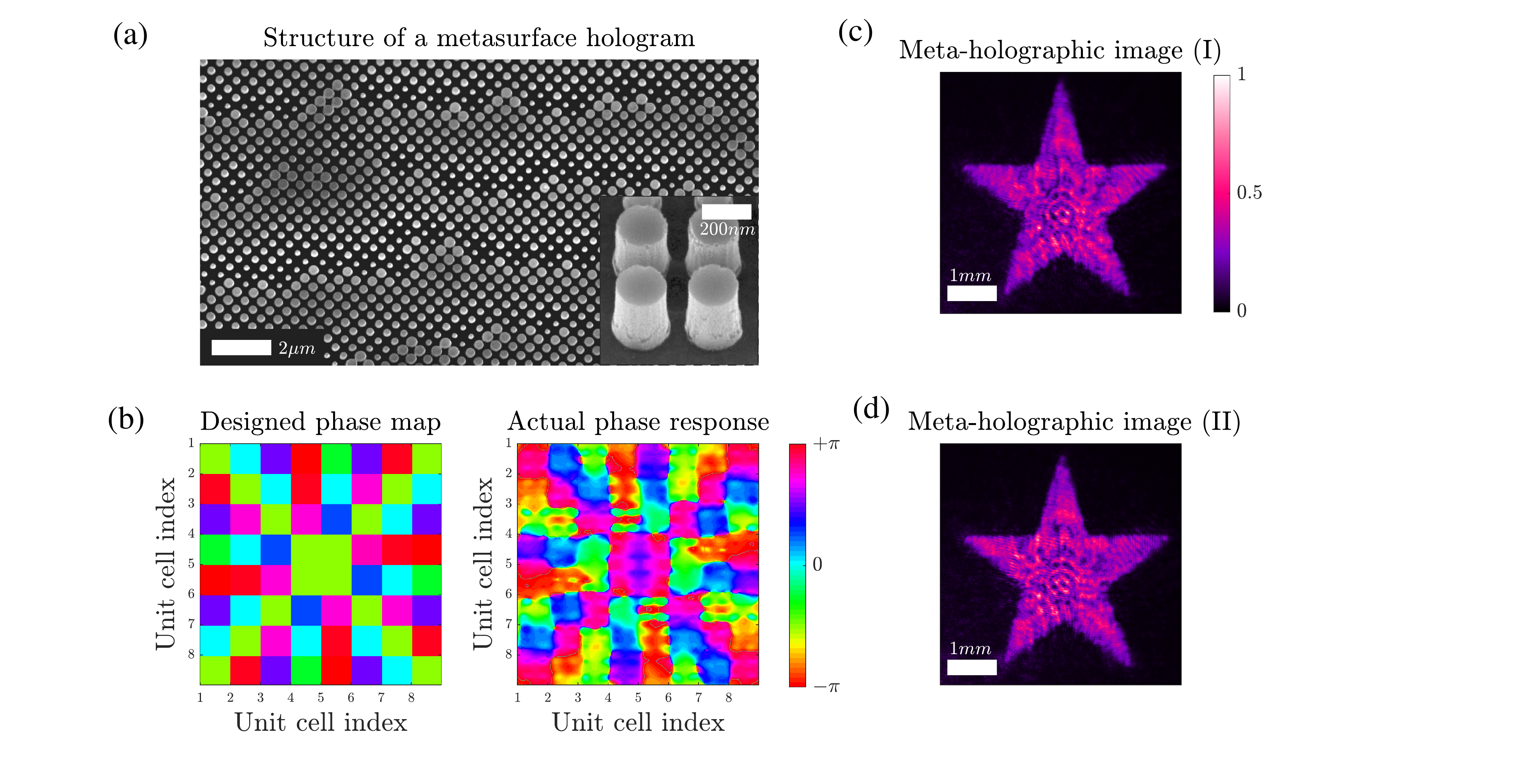}
    	%}
    	\caption{\textbf{Coherent artifacts of metasurface hologram}.
    		\textbf{(a)} Scanning electron microscope (SEM) image of a part of a meta-hologram comprised of $128 \times 128$ unit cells, each having $2 \times 2$ silicon nanopillars of the same diameter $D$. The spatial phase modulation is achieved by varying $D$ from 142 nm to 366 nm. 
    		\textbf{Inset:} Magnified view of the silicon nanopillars revealing surface roughness which causes unwanted scattering and interference of light. 
    		\textbf{(b)} \textbf{Left:} Designed phase $\phi_H$ map of a small meta-hologram comprised of $8 \times 8$ unit cells, based on the calculated phase response of individual nanopillars  with different diameters. 
    		\textbf{Right:} Actual phase response $\phi_A$ from a numerical simulation of the entire metasurface showing a significant deviation from the designed one due to near-field interactions among neighboring nanopillars. 
    		\textbf{(c,d)} Holographic images of a star object generated by two fabricated meta-holograms with the same design shown in (a). Their intensity fluctuations are nearly identical, indicating that the fluctuations result mainly from deterministic interactions among meta-atoms. Optical vortices are already eliminated from the computer-generated hologram. The illumination source is a monochromatic laser at wavelength $\lambda$~=~1064~nm, which has a high spatial and temporal coherence.}
    	\label{fig:pstatement}
    \end{figure}

    We design and fabricate a metasurface hologram as shown in Fig.~\ref{fig:pstatement}(a). It is made of Silicon nanopillars on top of a glass substrate (see \suppref{Methods} for the design and fabrication processes). There are $128 \times 128$ unit cells (pixels) in the meta-hologram, each consisting of $2 \times 2$ nanopillars of the same diameter $D$. The phase modulation of the meta-hologram is achieved via the resonant scattering of the individual nanopillars. By tuning the pillar diameter $D$, the scattering resonance frequency is varied and the phase response $\phi$ at the illumination wavelength is changed. In the meta-hologram design, $\phi(D)$ is calculated for a single pillar with periodic boundary conditions. To create a holographic image in the far-field, the near-field phase pattern $\phi_{H}$ is computed with an iterative phase retrieval (IPR) algorithm (see \suppref{Methods} for details). Then, the inverse mapping function $D(\phi)$ sets the nanopillar diameter $D$ in every pixel (unit cell) for the designed phase modulation $\phi_{H}$.
    
    Since the nanopillars are densely packed, the near-field interactions among neighboring nanopillars are significant. In the hologram design, the periodic boundary conditions used in the phase calculation correspond to the assumption that all neighboring pillars have an identical diameter $D$. This is not true in reality, as $D$ varies from one unit cell to the next. The near-field interactions between nanopillars with different diameters are different from the ones with the same diameter. Such difference causes the actual phase response ($\phi_A$) to deviate from the designed one ($\phi_H$). To illustrate this effect, we numerically simulate the actual phase modulation $\phi_A$ of a small meta-hologram with $8 \times 8$ meta-atoms. The phase modulation in Fig.~\ref{fig:pstatement}(b) is notably different from the designed pattern $\phi_H$. Such difference leads to a significant distortion of the holographic image, as observed experimentally in Fig.~\ref{fig:pstatement}(c). The seemingly-random intensity fluctuation is reproduced with another fabricated meta-hologram of identical design in Fig.~\ref{fig:pstatement}(d). The magnified images are available in the \suppref{Supplementary Materials}. Therefore, the artifacts are primarily due to deterministic cross-talk among the meta-atoms. 

    Such cross-talk is very difficult to correct, because a single meta-hologram is comprised of an enormous number of meta-atoms, e.g., the meta-hologram in Fig.~\ref{fig:pstatement}(a) contains $16\;384$ unit cells and $65\;536$ nanopillars. To accurately account for the interactions among meta-atoms of varying size, the phase response of the entire metasurface has to be calculated, a task which is computationally demanding. Any iterative optimization of the hologram configuration requires simulating the entire metasurface repetitively, which is beyond standard computing capabilities.
    In addition to the cross-talk of meta-atoms due to their near-field interactions, there are two more sources for meta-hologram artifacts. Due to the subwavelength size of the Silicon pillars, structural defects are introduced unintentionally during the fabrication of meta-hologram, as seen in the inset of Fig.~\ref{fig:pstatement}(a). Such defects induce unwanted light scattering and interference, producing additional artifacts in the holographic image. Furthermore, optical vortices are generated in the holographic image due to the creation of phase dislocations in the design of digital hologram. Such defects can be eliminated by incorporating complementary algorithms in the IPR \cite{Seldowitz87, wyrowski1989speckle, Feldman1989IterativeEO, senthilkumaran2005vortex, Maycock07} (see \suppref{Methods}). Alternatively, they are eliminated by full-field (amplitude and phase) modulation with a meta-hologram \cite{lee2018complete, overvig2019dielectric}.   

\newpage

\section{Degenerate cavity laser with tunable spatial coherence} \label{sec:laser}

    To remove coherent artifacts, we adjust the coherence properties of the illumination source. Since our meta-hologram is designed for a target frequency and is highly dispersive, lowering the temporal coherence by increasing the spectral bandwidth of illumination will reduce the hologram efficiency and degrade the image quality. Instead, we adjust the spatial coherence of illumination with a degenerate cavity laser (DCL).    
    
\begin{figure}[htbp]
        \centering
        %\fbox{
        \includegraphics[clip, trim = 6.5cm 8.6cm 1.4cm 1.3cm, width=1.0\linewidth]{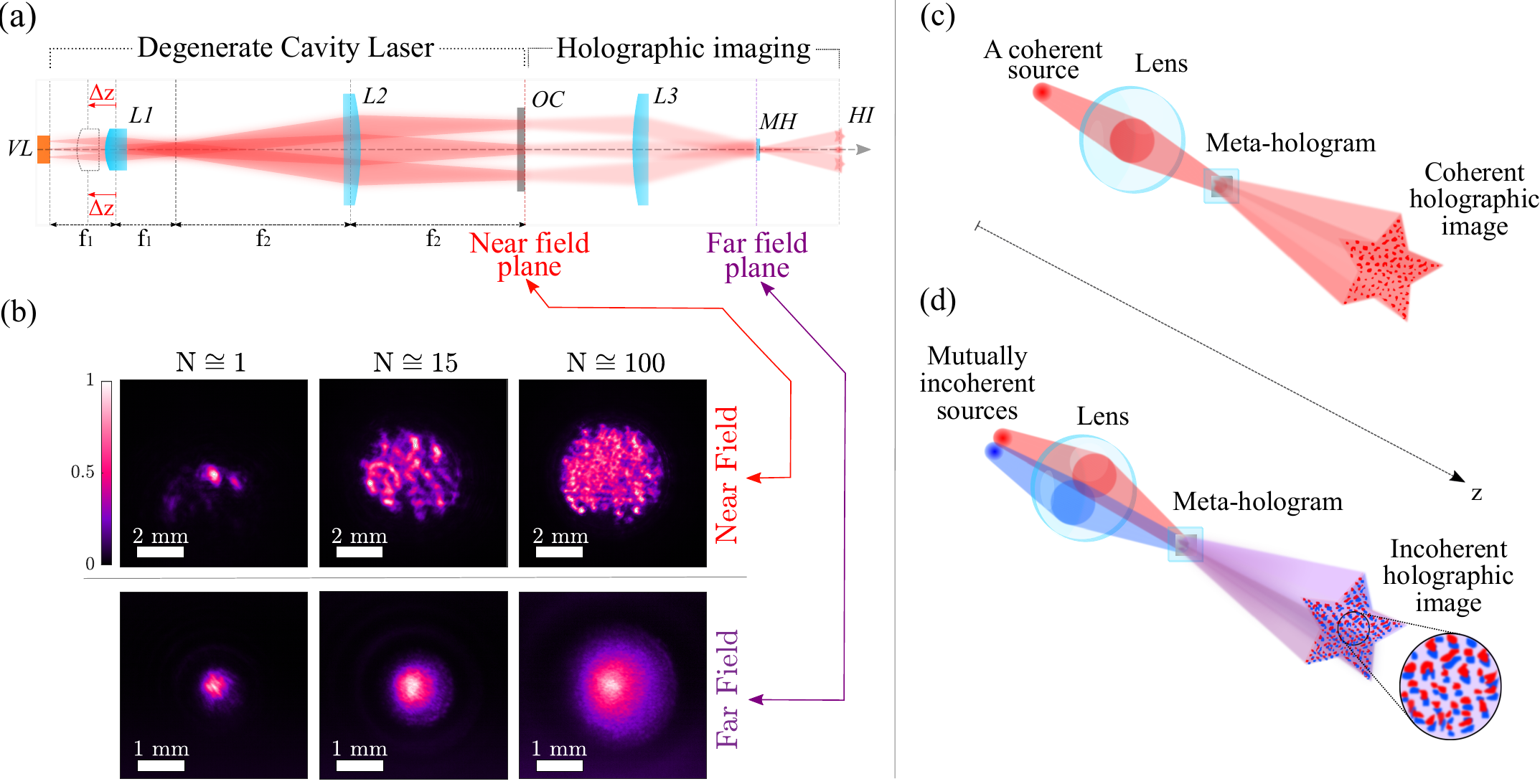}
        %}
        \caption{\textbf{Tuning the spatial coherence with a degenerate cavity laser (DCL).}        
        \textbf{(a)} Schematic of the DCL comprising of a vertical external cavity surface emitting laser (VECSEL) module (VL), two lenses (L1, L2) and an output coupler (OC). The far-field DCL emission is projected by imaging optics (L3) onto the meta-hologram (MH), which creates a holographic image (HI) at its far-field.
        L1 is gradually translated along the cavity axis to break the cavity degeneracy condition, so that the number of transverse lasing modes $N$ decreases and the spatial coherence of the total emission increases. 
        \textbf{(b)} Near-field \textbf{(top row)} and far-field \textbf{(bottom row)} intensity patterns of the total emission with a varying number $N$ of transverse lasing modes. With increasing $N$, there are more bright spots in the near-field (each corresponding to an independent lasing mode), and the far-field pattern becomes larger.
        \textbf{(c)} Schematic of a single lasing mode illuminating the meta-hologram, creating intensity fluctuations due to coherent artifacts. 
        \textbf{(d)} Schematic of two mutually incoherent lasing modes illuminating the meta-hologram with different angles, creating laterally shifted and mutually incoherent holographic images. An incoherent (intensity) sum of the two images reduces the intensity fluctuations.}
        \label{fig:VECSELNFandFF}
    \end{figure}

    The DCL has a self-imaging configuration \cite{Arnaud69}, as shown in Fig.~\ref{fig:VECSELNFandFF}(a). Since many transverse modes have a nearly degenerate quality factor, they can lase simultaneously and independently to reduce the spatial coherence of the emission. By tuning the cavity away from the degenerate condition (see \suppref{Methods}), the number of transverse lasing modes decreases, and the degree of spatial coherence increases.  
    
    Figure~\ref{fig:VECSELNFandFF}(b) shows the near-field (top row) and far-field (bottom row) intensity patterns of the laser emission. The near-field patterns at the DCL output coupler consist of bright spots, each corresponding to a transverse lasing mode. As the cavity approaches the degenerate condition, the number of spots (modes) increases. The diffracted beams from neighboring spots do not interfere, indicating that the modes are mutually incoherent. The number of independent lasing modes $N$ is estimated from the intensity contrast of a speckle pattern generated by a static diffuser placed outside the laser cavity (see \suppref{Methods} and \suppref{Supplementary Materials}). As $N$ decreases from $\sim 300$ to $\sim 1$, the emission power is merely reduced by $40\%$ from 108~mW to 64~mW (see \suppref{Supplementary Materials}). Furthermore, the emission spectrum remains approximately the same with a full-width-at-half-maximum (FWHM) of about 3 nm, indicating that the temporal degree of coherence does not change (see \suppref{Supplementary Materials}). This effectively avoids the influence of chromatic aberration.

    In contrast to the spotted near-field pattern, the laser emission exhibits a smooth profile at the far-field. It is composed of an incoherent superposition of Gaussian beams propagating in slightly different directions from individual spots at the near-field. The smooth intensity distribution ensures a homogeneous illumination of the meta-hologram which is placed at the far-field of the DCL. 
    
    The holographic image is created in the far-field of the meta-hologram. In the case of coherent illumination, the emission from a single transverse lasing mode illuminates the meta-hologram, as sketched in Fig.~\ref{fig:VECSELNFandFF}(c). With partially coherent illumination in Fig.~\ref{fig:VECSELNFandFF}(d), mutually incoherent lasing modes illuminate the meta-hologram with different angles of incidence and generate holographic images that are laterally shifted at the far-field. The number of distinct images is given by the effective number of independent spatial modes $N_E$ in illumination, which is equal to the ratio of the area of the hologram to the coherence area of the illuminating light (see \suppref{Methods}). An intensity sum of $N_E$ images will average out the intensity fluctuations due to coherent artifacts. However, the averaging also blurs the image and impair the spatial resolution. Hence, the degree of spatial coherence must be optimized to suppress coherent artifacts without significantly degrading the image resolution.    
  
\newpage

\section{Suppression of holographic artifacts} \label{sec:results}
    
\begin{figure}[htbp]
        \centering
        %\fbox{
        \includegraphics[clip, trim = 1.5cm 3.5cm 1.5cm 0cm, width=1\linewidth]{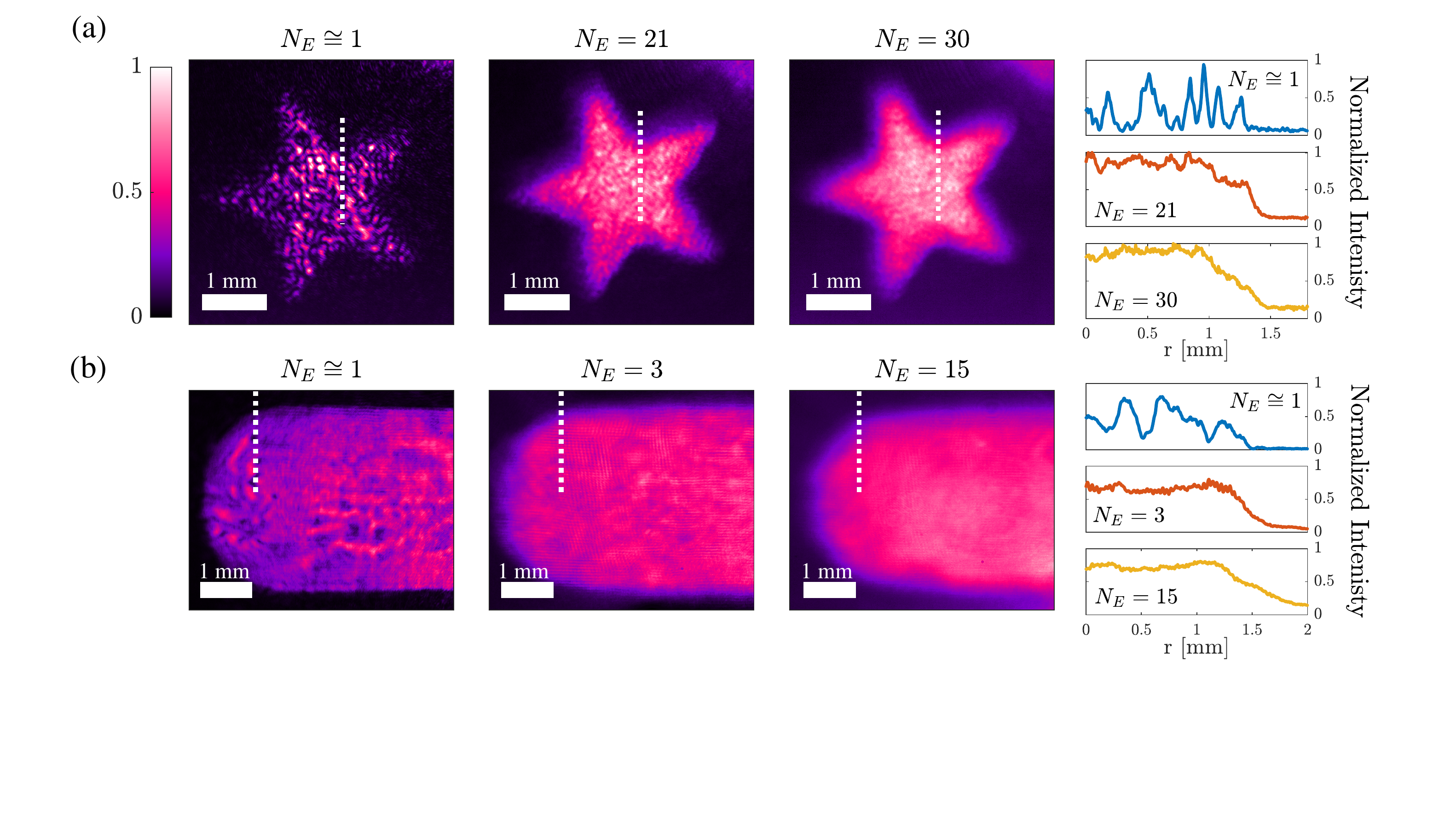}
        %}
        \caption{\textbf{Meta-holographic images under varying spatial coherence DCL illumination.} 
        The meta-holograms are based on the resonant phase modulation in \textbf{(a)} and the geometric Pancharatnam-Berry (PB) phase in \textbf{(b)}. The coherent artifacts seen with high spatial coherence illumination \textbf{(first column)} gradually disappear, as the spatial coherence is lowered by increasing the effective number of spatial modes $N_E$ that illuminate the meta-hologram  \textbf{(second column)}. Further decrease of the spatial coherence (increase of $N_E$) notably blurs the image and reduces the edge sharpness \textbf{(third column)}. The fourth column shows the 1D intensity profile through a cut of the holographic images marked by white dotted line in the first three columns. 
        }
        \label{fig:stars}
    \end{figure}

    To demonstrate the capability of our method in eliminating all sorts of coherent artifacts, we design one set of meta-holograms with the standard Gerchberg-Saxton IPR algorithm \cite{gerchberg1972practical}. The holographic images of this set contain many dark spots due to phase dislocations (optical vortices). The top row of Fig.~\ref{fig:stars} shows the holographic images, taken with varying degrees of spatial coherence of the DCL illumination. $N_E$ is the effective number of spatial modes that illuminate the meta-hologram and generate laterally shifted holographic images. When the spatial coherence is high ($N_E \cong 1$), the holographic image is full of coherent artifacts generated by near-field meta-atom interactions, fabrication defects and phase dislocations. Lowering the spatial coherence by increasing $N_E$ to $21$ suppresses the intensity fluctuations, resulting in a nearly uniform holographic image. A further increase of $N_E$ to $30$, however, results in a notable reduction of the edge sharpness, as seen in the 1D intensity profile across an edge of the star image in Fig.~\ref{fig:stars}. 
    
    We also test our method with meta-holograms free of phase dislocations. Despite the absence of optical vortices in the holographic images, intensity fluctuations are still significant, as seen in Fig.~\ref{fig:pstatement}(c,d). These artifacts can be effectively removed by optimizing the spatial coherence of the DCL illumination (see \suppref{Supplementary Materials}). 
    
    In addition to the resonant phase meta-holograms, our method is applicable to geometric Pancharatnam-Berry (PB) phase meta-holograms \cite{Bomzon02,chen2016review,cohen2019g}. These holograms are also cleared of phase dislocations originating from the phase encoding process. A scanning electron microscope image of the fabricated hologram is presented in the \suppref{Supplementary Materials}. The bottom row of Fig.~\ref{fig:stars} shows the holographic images recorded with the DCL illumination. Again, we observe intensity fluctuations under high spatial coherence illumination. In the absence of optical vortices, the artifacts result mainly from near-field interactions of meta-atoms. The intensity fluctuations are not as strong as those with optical vortices, thus a small decrease of the spatial coherence is sufficient to make the image smooth. However, the edges get blurred with a further reduction of the spatial coherence. The results of both types of meta-holograms confirm that the fine-tuning of the DCL spatial coherence is critical in achieving an optimal illumination condition where the holographic image is free of coherent artifacts and remains relatively sharp. 
    
\section{Optimal degree of spatial coherence}
    
\begin{figure}[htbp]
        \centering
        %\fbox{
        \includegraphics[clip, trim = 1.5cm 0cm 2cm 8cm, width=0.95\linewidth]{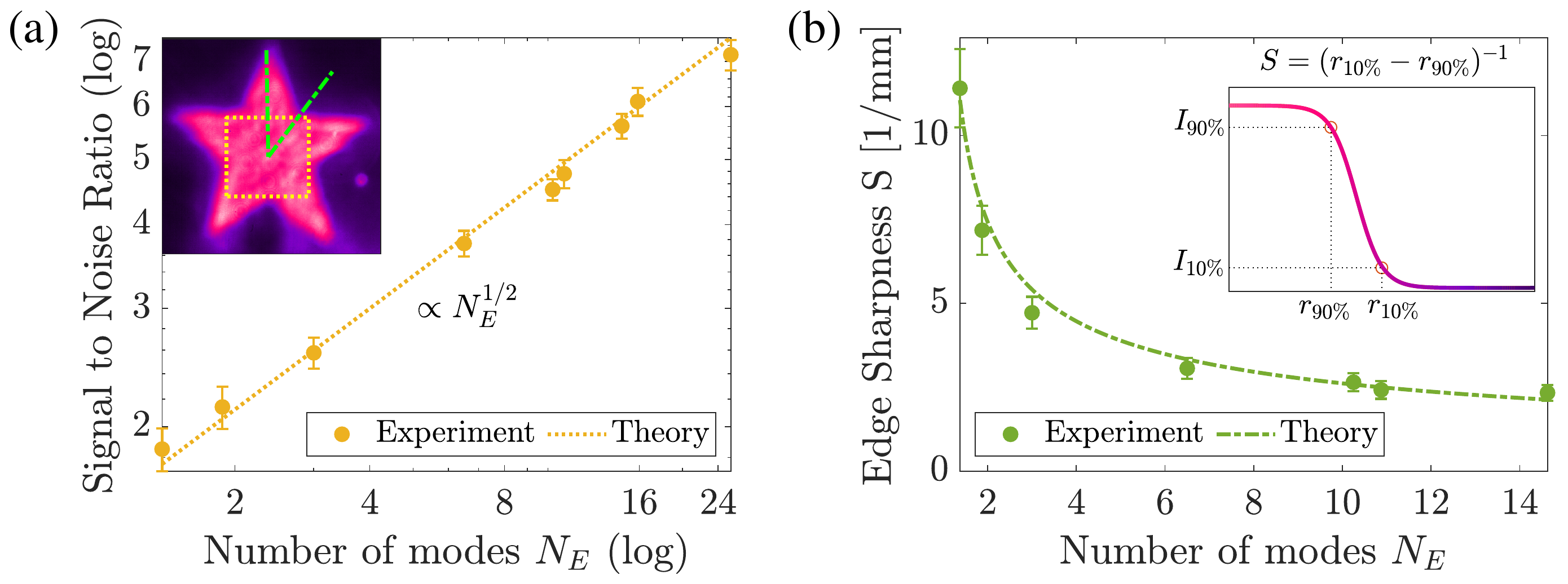}
        %}
        \caption{\textbf{Quantitative assessment of a meta-holographic image quality}. 
        \textbf{(a)} Intensity signal to noise ratio (SNR) of a star image (inset) within the central square (marked by a yellow dotted line in the inset) versus the number of spatial modes $N_E$ in illumination. Yellow circles are experimental data obtained by averaging the SNR over holographic images created by five meta-holograms. The error bars represent the standard deviation of the SNR. The data points follow the dotted yellow line with a slope of $1/2$ in logarithmic scales, indicating that the SNR scales as $\sqrt{N_E}$. 
        \textbf{(b)} Edge sharpness $S$ as a function of number of spatial modes $N_E$. Green circles represent experimentally measured $S$, averaged over multiple edges of five holographic images. The error bars represent the standard deviation of $S$. The dotted-dashed green line denotes the theoretical fit. 
        The inset illustrates how $S$ is extracted from the 1D intensity profile across an edge [marked by a dotted-dashed green line in the inset of (a)]. 
        }
        \label{fig:contrast}
    \end{figure}

    To quantitatively assess the holographic image quality, we evaluate the signal to noise ratio (SNR) and edge sharpness. Experimentally, we collect the data of five meta-holograms free of phase dislocations. The meta-holograms generate the same holographic image of a star, as shown in the inset of Fig.~\ref{fig:contrast}(a). The SNR is defined as $\langle I \rangle /\sigma$, where $\langle I \rangle$ is the average intensity within the central square marked in the inset of Fig.~\ref{fig:contrast}(a), and $\sigma$ is the standard deviation of the intensity fluctuation in this region. We average the SNR over five meta-holographic images, and plot its value versus the effective number of spatial modes $N_E$. As seen in Fig.~\ref{fig:contrast}(a), the SNR increases monotonically with $N_E$. In logarithmic scales, the data points follow a straight line of slope $1/2$, indicating that the SNR scales as $\sqrt{N_E}$.

    The edge sharpness is estimated from several 1D intensity profiles of the holographic image across different edges, as marked by green dotted-dashed lines in the inset of Fig.~\ref{fig:contrast}(a). The sharpness is evaluated by estimating the slope of the edge, between two intensity points corresponding to $10\%$ and $90\%$ of the maximum intensity, $S=1/\left(r_{10\%}-r_{90\%}\right)$ [see the inset of Fig.~\ref{fig:contrast}(b)]. Figure~\ref{fig:contrast}(b) shows the edge sharpness $S$ averaged over multiple edges of five holographic images. As the effective number of spatial modes $N_E$ grows, $S$ drops monotonically. 
    Next we model the dependence of $S$ on $N_E$. For $N_E=1$, the edge intensity profile is given by convolution of the ideal step function with the point spread function (PSF) of the holographic imaging setup. The width $w_P$ of the PSF determines the spatial resolution, and is inversely proportional to the lateral dimension of the meta-hologram. For $N_E>1$, $N_E$ laterally shifted holographic images are created, and an incoherent summation of all images broadens the edge intensity profile. The broadening depends on the width of the DCL near-field emission pattern [Fig. \ref{fig:VECSELNFandFF}(b)], and is proportional to $\sqrt{N_E-1}$. As a result of convolution, the sharpness scales as $S(N_E) \propto \left[{w_P^2+C_1 \cdot (N_E-1) }\right]^{-1/2}$, where $C_1$ is a scaling constant relating the lateral shift of a holographic image to the tilt of incident angle of an illuminating beam (see \suppref{Supplementary Materials} for a complete derivation). By fitting the experimental data with the above expression of $S(N_E)$, we obtain the green curve in Fig.~\ref{fig:contrast}(b), which agrees well with the measured dependence of $S$ on $N_E$.

\newpage

\begin{figure}[htbp]
        \centering
        %\fbox{
        \includegraphics[clip, trim = 0cm 0cm 0cm 0cm, width=0.9\linewidth]{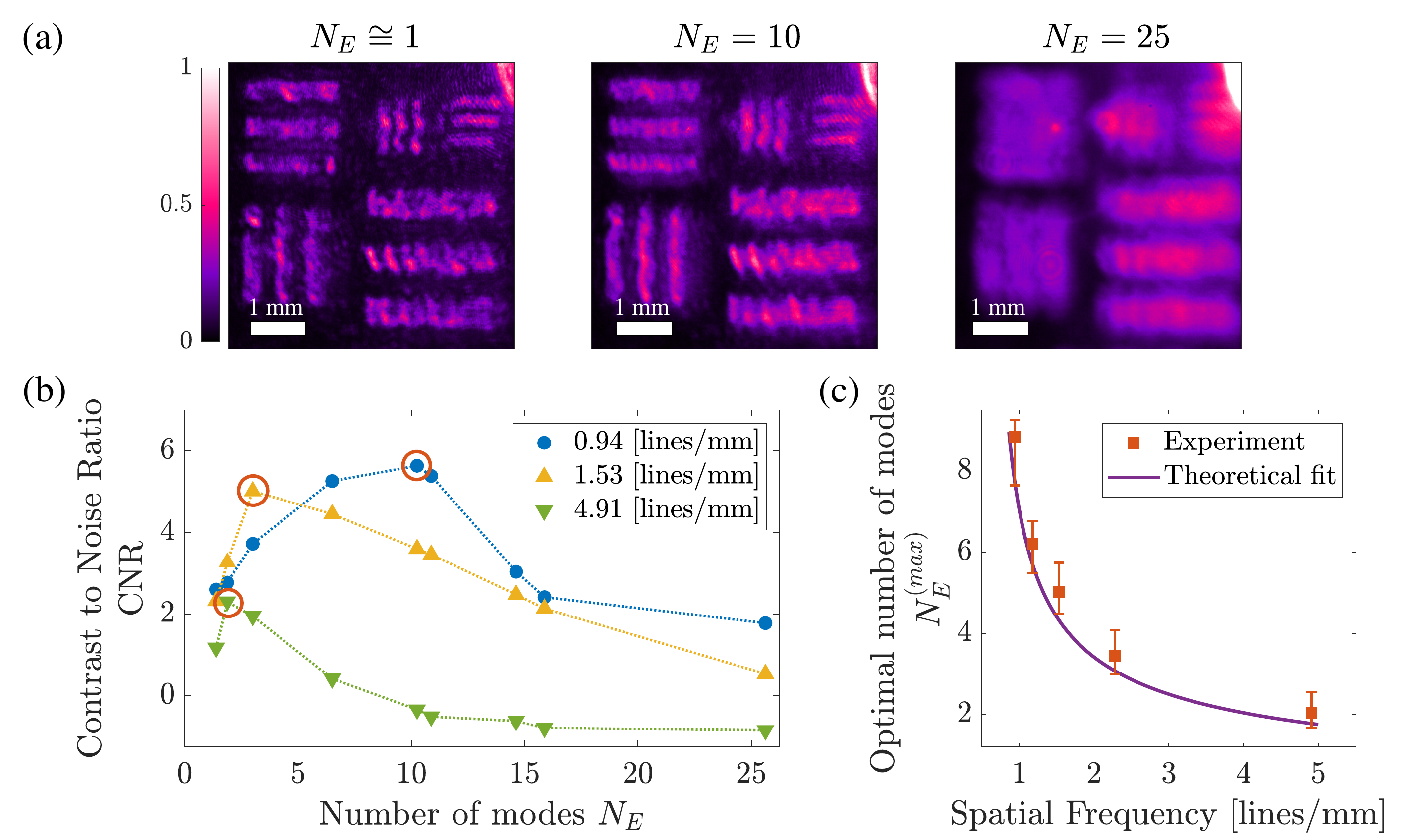}
        %}
        \caption{\textbf{Optimal degree of spatial coherence}. 
        \textbf{(a)} Three holographic images of USAF test charts obtained with the same meta-hologram and illuminated by the DCL with varying degrees of spatial coherence. As the effective numbers of spatial modes $N_E$ increases, the spatial coherence of the illumination decreases and the coherent artifacts are suppressed, however, the image resolution is impaired. 
        \textbf{(b)} Contrast to Noise ratio (CNR) versus the effective number of modes $N_E$ for three spatial frequencies in the test charts. The CNR first increases with $N_E$, reaches a maximum (circled in red) and then decreases. The maximum of the CNR shifts to lower $N_E$ for higher spatial frequencies (smaller feature size). 
        \textbf{(c)} The optimal number of spatial modes $N_E^{(max)}$ at the maximal CNR decreases, as the spatial frequency (inverse of spatial resolution) increases. The purple curve is the theoretical fit.
        }
        \label{fig:cnr}
    \end{figure}

    Finally, we identify the optimal degree of spatial coherence for illuminating a meta-hologram and find its dependence on the image resolution. To this end, we fabricate another set of meta-holograms that produce holographic images of a USAF resolution test chart. Figure~\ref{fig:cnr}(a) shows three holographic images with different degrees of spatial coherence illumination. The image quality is assessed by the contrast to noise ratio (CNR), which is defined as
    \begin{equation}  
    CNR \equiv \frac{\left<I_S\right>-\left<I_B\right>}{\sigma_S} = \frac{1-\left<I_B\right>/\left<I_S\right>}{\sigma_S/\left<I_S\right>},
    \end{equation}    
    
    where $\left<I_S\right>$ and $\sigma_S$ denote the average intensity and the standard deviation of the intensity fluctuation in the bright region, respectively. $\left<I_B\right>$ is the average intensity of the dark background. The numerator of the CNR, $1-\left<I_B\right>/\left<I_S\right> $, gives the intensity contrast between bright and dark regions, while the denominator $\sigma_S /\left<I_S\right>$ characterizes the intensity fluctuation (noise) in the bright region. Overall, the CNR describes the resolvability of a feature of interest (bright) in a given background (dark).

    Figure~\ref{fig:cnr}(b) shows the measured CNR varying non-monotonically with the effective number of spatial modes $N_E$ for three different feature sizes. As $N_E$ increases, the CNR first grows and then drops. It reaches a maximal value at an intermediate $N_E$, indicating that there is an optimal degree of spatial coherence for illumination. When the feature size is large, the CNR reaches the maximum at a relatively large $N_E$. Since the intensity contrast (in the numerator of CNR) remains high for a relatively wide range of $N_E$, the CNR is determined mainly by the intensity fluctuation (in the denominator), which is smaller at larger $N_E$. As the feature size decreases, the maximum CNR shifts to a lower $N_E$. That is because, in resolving small features, the intensity contrast becomes more significant and is higher at smaller $N_E$ due to less blurring. However, the maximal value of the CNR is less than that for a large feature size, because the intensity fluctuations are stronger. Therefore, the optimal degree of spatial coherence increases with the image resolution.
       
    Figure~\ref{fig:cnr}(c) shows the optimal number of independent spatial modes $N_E^{(max)}$ (number of laterally shifted holographic images) required to reach the maximum CNR as a function of the spatial frequency (inverse of spatial resolution) in the USAF test chart. As the spatial frequency increases, the feature size decreases, and $N_E^{(max)}$ drops. Our theoretical modeling of the CNR (see \suppref{Supplementary Materials}) predicts $N_E^{(max)}$ (purple line) in good agreement with the experimental data.

\newpage

\section{Discussion and Conclusion} \label{sec:conclusions}

    In this work, we present an efficient method to suppress coherent artifacts of metasurface holograms. The majority of the meta-hologram image artifacts originates from near-field interactions of closely-packed meta-atoms. Therefore, the spatial intensity fluctuations are deterministic and extremely difficult to eliminate in the design of a meta-hologram. By optimizing the illuminating light spatial coherence, we suppress artifacts created by cross-talk as well as by phase dislocations and fabrication defects. 
    
    Our illumination source is a degenerate cavity laser (DCL). It supports simultaneous, independent lasing in many transverse modes with nearly degenerate loss, producing emission with a low spatial coherence. By misaligning the cavity to lift the loss degeneracy of transverse modes, the number of lasing modes decreases and the spatial coherence of emission increases. The continuous tuning of the DCL spatial coherence is energy efficient, and does not introduce a significant power variation. The spectral width of the DCL emission (degree of temporal coherence) remains constant during the spatial coherence tuning, which is important for meta-holograms with strong dispersion. Even with multimode operation, the spectral radiance (photon degeneracy number) of our DCL exceeds that of a superluminescent diode (SLD) by one order of magnitude and a light emitting diode (LED) by six orders of magnitude (see \suppref{Supplementary Materials}).   
    
    As the spatial coherence of illumination decreases, the signal to noise ratio of a meta-holographic image improves, but the edge sharpness drops. The precise tuning of the DCL spatial coherence is essential in finding an optimal degree of spatial coherence to effectively suppress coherent artifacts while maintaining a relatively high sharpness. The optimal degree of spatial coherence, at which the contrast to noise ratio reaches the maximum, depends on the image resolution. A finer spatial resolution requires higher spatial coherence.   
    
    In summary, our scheme can quickly remove all coherent artifacts created by different types of meta-holograms. It paves the way for the applications of meta-holograms in dynamic display, augmented reality, optical storage, beam multiplexing, nonlinear holography and optical manipulation. 

\section{Methods}

    \subsection{Digital meta-hologram design}
        
        We design two types of metasurface holograms. In the first one, the phase modulation is obtained via resonant scattering of silicon nanopillars with varying diameter. The second type is based on a geometric Pancharatnam-Berry (PB) phase modulation induced by silicon nanofins of different in-plane orientation.      
        Using a commercially available finite element method  (FEM) solver (COMSOL Multiphysics), we calculate the phase response $\phi$ of a single nanopillar with diameter $D$ and a nanofin with orientation angle $\theta$. Periodic boundary conditions are applied, thus neighboring nanopillars (nanofins) are assumed to have an identical diameter $D$ ($\theta$). By varying $D$ ($\theta$), we obtain the mapping  $\phi(D)$  [$\phi(\theta)$]. 
        We use an iterative phase retrieval (IPR) algorithm to find the near-field phase modulation creating a far-field holographic image. The standard method based on the Gerchberg-Saxton (GS) algorithm \cite{gerchberg1972practical}  generates optical vortices (phase dislocations) in the holographic image. To remove such artifacts, the GS algorithm is modified with an initial spherical phase front \cite{pang2019speckle} and followed by a simulated annealing (SA) routine \cite{Seldowitz87,Feldman1989IterativeEO}. 
        Finally, we encode the near-field phase profile $\phi_H$, pixel by pixel, in a metasurface comprising of $128\times128$ unit cells. Each unit cell (pixel) contains $2 \times 2$ nanopillars of the same size  (nanofins of the same orientation). The nanopillar diameter $D$ (nanofin angle $\theta$) in every unit cell is chosen from the inverse mapping $D(\phi)$ ($\theta(\phi)$). For the nanopillar hologram, $2 \pi$ phase modulation is achieved by varying $D$ from 142 nm to 366 nm (\suppref{see Supplementary Materials}). In the nanofin hologram, all nanofins an have identical size but varying orientation. Each fin has a length of 393 nm, a width of 82 nm, and a thickness of 600 nm. The geometric phase $\phi$ is dictated by the in-plane orientation angle $\theta$:  $\phi = \pm 2 \theta$, where $\pm$ signs correspond to left- and right-circular polarizations of incident light. With $\theta$ varying from 0 to $\pi$, the phase $\phi$ modulation covers a $2 \pi$ range. When illuminated by the linear polarized emission from the DCL, the nanofin hologram generates two images of left and right circular polarizations at different far-field locations.
        While the design of the meta-hologram phase profile $\phi_H$ is done by tiling pixels with the pre-calculated phase response of individual meta-atoms, the actual phase response $\phi_A$ of a small meta-hologram with $8\times8$ unit cells is calculated by the FEM with absorbing boundary conditions, and shown in the right panel of Fig. \ref{fig:pstatement}(b).
        
    \subsection{Meta-surface fabrication}
    
        The silicon (Si) metasurfaces are fabricated with electron-beam lithography and reactive ion etching. A 600-nm-thick amorphous Si film is deposited on a glass substrate using an electron beam evaporator (SKE$\_$A$\_$75). Then, a 100-nm-thick electron-beam resist (MicroChem PMMA [polymethyl methacrylate] A2) is spin-coated onto the Si film and patterned with the electron beam writer (Raith E-line, 30kV). After development in a MIBK ${\&}$ IPA (1:3) solution, 15-nm-thick Chromium (Cr) is deposited on the sample using electron beam evaporation (SKE$\_$A$\_$75) and the inverse nano-pattern is transferred to the Cr layer by a lift-off process in a remover PG (Micro Chem). By etching the Si with Sulfur hexafluoride (SF$_6$) and Fluoroform (CHF$_3$) in a flow of 5 sccm and 50 sccm, respectively, the nano-patterns are transferred to the Si membrane. Finally, the Cr mask is removed by immersing the samples in a Cr etchant solution (Aldrich Chemistry) for 30 minutes.
        
    \subsection{Spatial coherence tuning of the DCL}\label{sec:cohTuning}
    
        The coherence level of the DCL is tuned by translating a lens [L1 in Fig.~\ref{fig:VECSELNFandFF}(a)] inside the cavity along the longitudinal axis. The lens is mounted on a mechanical translation stage with micrometer resolution (Thorlabs MBT616D). When the lens L1 is accurately positioned ($\Delta z$ = 0 $\mu$m) to satisfy the self-imaging condition, lasing occurs in many transverse modes with nearly degenerate loss, and the total emission has a low spatial coherence. Moving the lens L1 from $\Delta z = 0$ $\mu$m breaks the degeneracy condition and reduces the number of transverse lasing modes. At $\Delta z$ = 300 $\mu$m, nearly all transverse modes stop lasing except one and the spatial coherence of emission is high. By gradually changing $\Delta z$ from 0 $\mu$m to 300 $\mu$m, we can continuously vary the number of transverse lasing modes, and accurately tune the degree of spatial coherence from low to high.
        
    \subsection{Characterization of spatial coherence}
    
        To measure the number of independent transverse lasing modes $N$ in the DCL, we direct the lasing emission to a spatial coherence measurement setup. It consists of two lens (L3, L4) with identical focal length $f$, which are arranged in a $4f$ configuration. A ground glass diffuser (Thorlabs DG10-600) is placed in between L3 and L4 at the mutual focal plane (see \suppref{Supplementary Materials} for a schematic and more details). The speckle pattern generated by the diffuser is measured by a CCD camera at the back focal plane of the second lens L4. The intensity contrast $C$ of the speckle pattern is defined as $C \equiv \sigma_I / \langle{I\rangle}$, where $\langle I \rangle $ is the mean intensity and $\sigma_I$ is the standard deviation of intensity fluctuation. $C$ is related to the number of independent transverse lasing modes $N$ by $C = 1/\sqrt{N}$. For different detunings ($\Delta z$) of the DCL, the number of transverse lasing modes is estimated from the measured speckle contrast: $N=1/C^2$.
    
        In order to validate the mutual incoherence of the bright isolated spots in the near-field emission pattern of DCL [Fig. \ref{fig:VECSELNFandFF}(d)], we measure the spatial pattern of the output beam as it propagates away from the DCL. The diffraction of emission from individual spots causes a spatial overlap of neighboring ones, but no interference fringes are observed in the time-integrated emission pattern, indicating these lasing spots are mutually incoherent.
         
        In the holographic imaging setup, the diameter $L_I$ of the illuminating beam is larger than the lateral dimension $L_H$ of a meta-hologram to ensure a uniform intensity illumination. The angular width of the illuminating beam $\Delta \theta_I$ is inversely proportional to the spatial coherence length $L_C$: $\Delta \theta_C \propto 1/ L_C$. The ratio between the illuminating beam area $A_I \propto {L_I}^2$ and the coherence area $A_C \propto {L_C}^2$ gives the number of independent spatial modes: 
        \begin{equation}
            N = \frac{A_I}{A_C} = \left( \frac{L_I}{L_C} \right)^2.
        \end{equation}  
        Since the illuminating beam area is larger than the meta-hologram area $A_H = {L_H}^2$, the effective number of modes $N_E$ within the meta-hologram is smaller than $N$. The effective number of modes $N_E$ interacting with the meta-hologram is given by the ratio of the meta-hologram area and the illumination coherence area:
        \begin{equation}
            N_E = \frac{A_H}{A_C} = \left( \frac{L_H}{L_C} \right)^2 \propto \frac{\Delta \theta_I}{\Delta \theta_H},
        \end{equation}      
        where $\Delta \theta_H \propto 1/ L_H$ is the diffraction angle of the meta-hologram due to its finite size.
        The effective number of spatial modes in illumination $N_E$ gives the number of independent holographic images generated at the far-field. In our setup, the ratio $N/N_E = A_I / A_H$ is found to be approximately 8.
        
\bibliography{bibliography}

\section*{Acknowledgements}

    Y.E. would like to thank Nicholas Bender for fruitful discussions. The work done at Yale University is supported by the US Office of Naval Research (ONR) under Grant Nos. N00014-21-1-2026 and N00014-20-1-2197. Q.S., S.X. thank the financial support from National Key Research and Development Program of China (Grant No. 2018YFB2200403), National Natural Science Foundation of China (Grant Nos. 12025402, 11974092, 91850204, 11934012, and 61975041), Shenzhen Fundamental research projects (Grant Nos. JCYJ20180507184613841, JCYJ20180507183532343, and JCYJ20180306172041577)

\section*{Author contributions statement}

    Y.E. conducted experiments, analyzed data and performed theoretical analysis, G.Q. conducted experiments, designed the meta-holograms and produced simulation results, W.Y., Y.W. and S.X. fabricated the meta-holograms. H.Y. participated in discussions and provided suggestions, Q.S. and H.C. supervised the research project. All authors reviewed the manuscript.

\section*{Competing interests}

    The authors have no competing interests.

\end{document}